\documentclass{article}
\usepackage{spconf,amsmath,graphicx}
\usepackage{color}
\usepackage{multirow}
\usepackage{lipsum}  
\usepackage{booktabs,siunitx}
\usepackage{colortbl}
\usepackage{amssymb}
\usepackage{pifont}
\usepackage{amsfonts}
\usepackage{tabularx}
\usepackage{array, makecell}
\usepackage{graphicx}
\usepackage{xcolor}
\usepackage{kotex}
\usepackage{comment}
\usepackage{soul}
\usepackage{url}
\usepackage{amsmath}
\usepackage{kotex}
\usepackage{tabularray}

\title{Enriching Music Descriptions with a Finetuned-LLM and Metadata \\ for Text-to-Music Retrieval}
\name{$^\flat$SeungHeon Doh, $^\natural$Minhee Lee, $^\sharp$Dasaem Jeong, $^\flat$Juhan Nam} 
\address{$^\flat$Graduate School of Culture Technology, KAIST, South Korea \\ 
$^\natural$Department of Computer Science and Engineering, Sogang University, South Korea \\ 
$^\sharp$Department of Art \& Technology, Sogang University, South Korea
}
\begin{document}
\ninept
\maketitle
\begin{abstract}
Text-to-Music Retrieval, finding music based on a given natural language query, plays a pivotal role in content discovery within extensive music databases. To address this challenge, prior research has predominantly focused on a joint embedding of music audio and text, utilizing it to retrieve music tracks that exactly match descriptive queries related to musical attributes (i.e. genre, instrument) and contextual elements (i.e. mood, theme). However, users also articulate a need to explore music that shares similarities with their favorite tracks or artists, such as \textit{I need a similar track to Superstition by Stevie Wonder}. To address these concerns, this paper proposes an improved Text-to-Music Retrieval model, denoted as \textbf{TTMR++}, which utilizes rich text descriptions generated with a finetuned large language model and metadata. To accomplish this, we obtained various types of seed text from several existing music tag and caption datasets and a knowledge graph dataset of artists and tracks. The experimental results show the effectiveness of TTMR++ in comparison to state-of-the-art music-text joint embedding models through a comprehensive evaluation involving various musical text queries. \footnote{Our dataset and codes are available at \url{https://github.com/seungheondoh/music-text-representation-pp}}


\end{abstract}

\section{Introduction}
\label{sec:intro}

Text-to-music retrieval is the task of retrieving the music items that are most relevant to the natural language query~\cite{turnbull2008semantic}. To find music that matches user preferences within large music databases, users may use queries in the free-form text description. Previous works have focused on learning embedding functions that map music and text inputs into a joint embedding space, simplifying text-to-music retrieval to cross-modal nearest neighbor retrieval~\cite{choi2019zero, won2021multimodal, manco2022contrastive, huang2022mulan, doh2023toward, weck2023wikimute, wu2023clamp}.

Despite the exploration of various music-text joint embedding models, there are still several limitations in text-to-music retrieval systems when it comes to comprehending diverse textual user queries. A considerable body of research~\cite{choi2019zero, won2021multimodal, manco2022contrastive, doh2023toward} is dedicated to learning a joint embedding space that bridges the gap between music and content semantics. However, users may formulate queries that encompass not only text related to content semantics such as genre, mood, and theme but also queries involving metadata such as track, album, and artist entity names. In these cases, users may employ queries not only for exact matching (e.g., ``music from Stevie Wonder") but also for those related to relative similarity with a degree of tolerance (e.g., ``similar music to Stevie Wonder"). Some prior studies~\cite{huang2022mulan, wu2023clamp} have integrated metadata text (e.g., track title, artist, composer) into their training processes; however, they did not explicitly evaluate the metadata-based retrieval in their assessment. What further complicates this comprehensive evaluation is the absence of large-scale music-text paired datasets containing content descriptions, metadata, and knowledge of relative similarity.

\begin{figure}[!t]
\centering
\includegraphics[width=\linewidth]{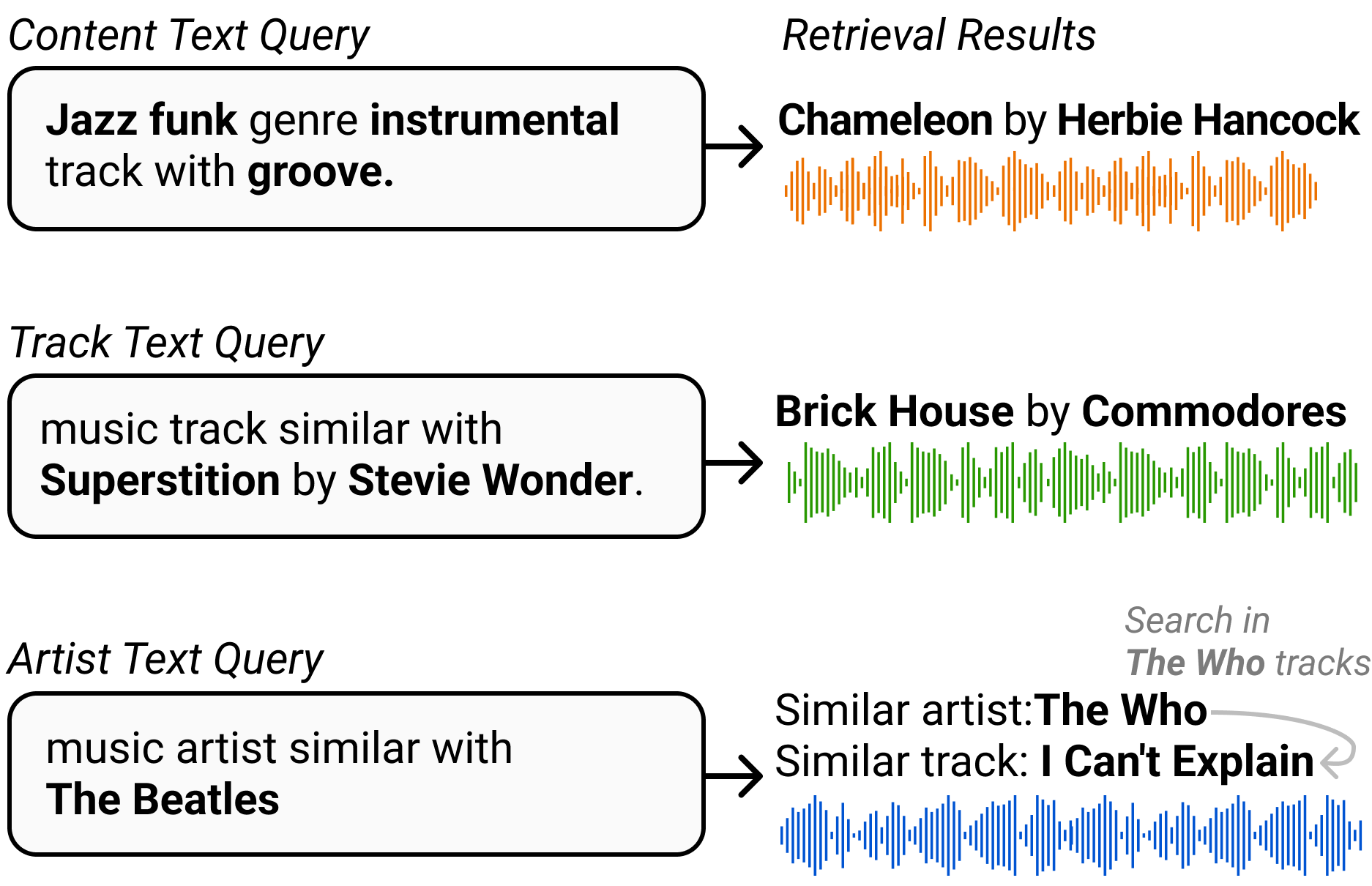}
\vspace{-6mm}
\caption{An illustration of text-to-music retrieval scenario with content, track, and artist queries.}
\vspace{-5mm}
\label{fig:teaser}
\end{figure}

One approach to handling the challenge of low-resource data involves training models on multiple datasets using a unified training framework. For instance, Gardner et al.~\cite{gardner2021mt3} proposed a sequence-to-sequence framework to concurrently train on various music transcription datasets. Wu et al.~\cite{wu2022large} also jointly trained the model on various audio-text datasets in a cross-modal dual encoder framework. Another method involves data augmentation through the utilization of a large language model (LLM). Doh et al.~\cite{doh2023lp} proposed a method for generating pseudo music captions from existing tagging datasets and an LLM (i.e., GPT3.5-175B+). Similarly, Mckee et al.~\cite{mckee2023language} generated music descriptions using a pre-trained music tagger and an LLM (i.e., BLOOM-176B). In the general audio domain, Mei et al.~\cite{mei2023wavcaps} applied an LLM-based pseudo-caption generation approach to various datasets, resulting in performance improvements in audio captioning, text-to-audio retrieval, and text-to-audio generation.

\begin{figure*}[!t]
\centering
\includegraphics[width=\linewidth]{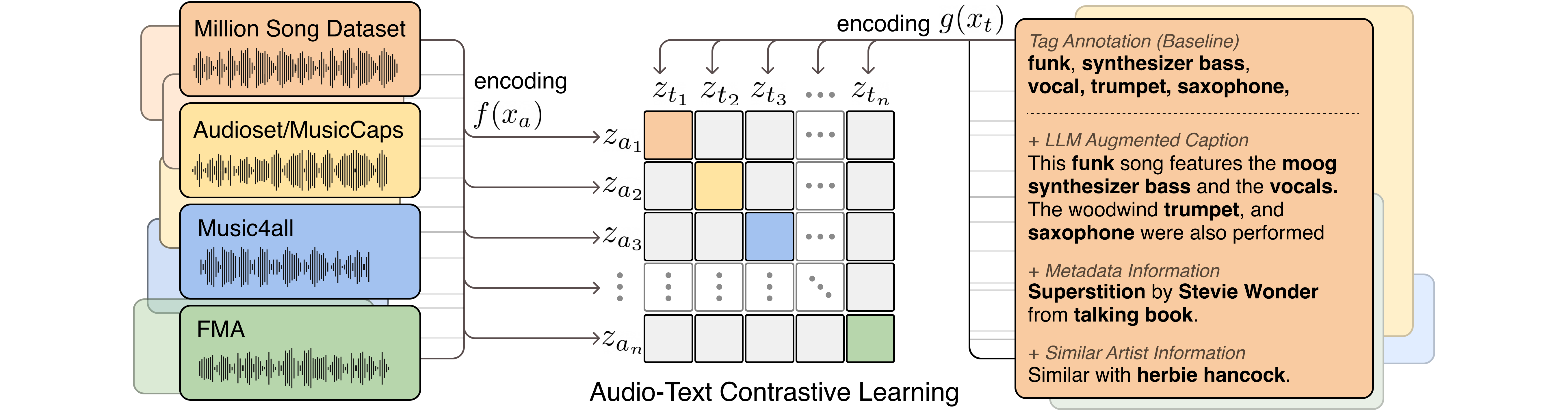}
\vspace{-5mm}
\caption{Music Audio and Text Contrastive Learning Framework.}
\vspace{-3mm}
\label{fig:main}
\end{figure*}

In this paper, we aim to design a music-text joint embedding model that comprehends content and metadata semantics, enabling not only exact matching with audio but also an understanding of relative similarity (illustrated in Figure~\ref{fig:teaser}). Our contributions for this purpose are as follows:

\vspace{1mm}
\noindent \textbf{Joint training framework:} We utilize five diverse music tagging datasets, encompassing various types of content-related information (genre, instrument, mood, theme) as well as metadata (track, album, artist name), which leads to the collection of a public music-text pair dataset comprising 1.3 million entries.

\vspace{1mm}
\noindent \textbf{Pseudo caption generation with a finetuned LLM:} We generate pseudo-captions using existing tag annotation datasets and an LLM. For this process, we fintuned LLaMA2~\cite{touvron2023llama} as an instruction-following model using \textsc{\{}instruction, tag input, and pseudo-caption output\textsc{\}} data generated by the GPT3.5.

\vspace{1mm}
\noindent \textbf{Music description from metadata knowledge graph:} We generate music descriptions for audio tracks through objective metadata associations and synthesize a track-artist similarity text by utilizing both artist similarity (artist$\leftrightarrow$artist) and metadata association (artist$\leftrightarrow$track).


\section{Text-to-Music Retrieval}
In this section, we introduce our improved Text-to-Music Retrieval (TTMR++) framework. At a high level, our model is presented with paired text and music audio samples and is trained to map these paired samples into a joint embedding space. This mapping aims to maximize the similarity between the vectors representing music audio and text, as depicted in Figure~\ref{fig:main}. In the subsequent explanations, we denote a musical audio sample as $x_{a}$ and a paired text sample as $x_{t}$. The functions $f(\cdot)$ represent an audio encoder, while $g(\cdot)$ symbolizes a text encoder. These encoders comprise a backbone model, a linear projection with $d$=128 units, and an $l_{2}$ normalization layer. The resulting embeddings for audio and text are represented as $z_{a} = f(x_{a})$ and $z_{t} = g(x_{t})$, respectively.

\subsection{Cross-Modal Dual Encoder with Contrastive Loss}
\label{sec:dual_encoder}
We use a cross-modal dual encoder model with the contrastive loss that has achieved outstanding results in cross-modal retrieval tasks~\cite{manco2022contrastive, huang2022mulan, doh2023toward, radford2021learning}. We adopt contrastive loss to reduce the distance between positive sample pairs while increasing the distance between negative sample pairs. During training, the audio and text encoders are jointly trained to maximize the similarity between $N$ positive pairs of (music, text) associations while minimizing the similarity for $N \times (N-1)$ negative pairs. This is known as the multi-modal version of InfoNCE loss~\cite{oord2018representation, radford2021learning} and is formulated as follows:

\vspace{-2mm}
\begin{equation}
\label{eq2}
\begin{aligned}
\mathcal{L}_{t \rightarrow a} = - \log \frac{\exp(z_{t,i} \cdot z_{a,i} / \tau)}{\sum^{N}_{j=1} \exp( z_{t,i} \cdot z_{a,j}) / \tau)} \\
\end{aligned}
\end{equation}
where $\tau$ is a learnable parameter. The loss function is designed as follows: $\mathcal{L}_{t \leftrightarrow a} = (\mathcal{L}_{t \rightarrow a} + \mathcal{L}_{a \rightarrow t})/2
$.

\subsection{Audio and Text Encoder}
We have selected modified version of ResNet-50~\cite{radford2021learning} as our audio encoder. With a convolutional layer and residual layers featuring bottleneck blocks, the ResNet-50 processes input log-mel spectrograms. Following this, attention pooling is applied to the outputs of the residual layers to capture global audio features. This attention pooling is implemented as a single layer of multi-head QKV attention, where the query is derived from average-pooled audio features, and keys and values correspond to sequential audio features. The resulting global audio features undergo layer normalization and linear projection, mapping them into the joint embedding space.

Our text encoder is RoBERTa~\cite{liu2019roberta}, a masked language model trained on a massive 160GB text corpus. The input text sequence is tokenized using a byte pair encoding (BPE) tokenizer, with a maximum sequence length of 128 tokens. Following tokenization, the BPE token embeddings are processed through 12 transformer blocks, each with a width of 768 units. Similar to \cite{huang2022mulan, doh2023toward}, we extract the output from the first position (i.e., the start-of-sentence token) as the feature representation of text, applying a layer normalization and a linear projection to map it into the joint embedding space.

\section{Enriching Music Descriptions}
\label{sec:semantic}

\subsection{Music Description from a Finetuned-LLM}
\label{sec:llm}
To generate high-quality music descriptions, we leverage the open-source LLM, LLaMA2-7B~\cite{touvron2023llama}, in conjunction with an existing tagging dataset. To achieve performance comparable to that of the closed-source LLM, GPT-3.5~\cite{ouyang2022training}, we employed an instruction fine-tuning approach~\cite{taori2023stanford}, transforming LLaMA2-7B into an instruction-following model. Starting from the LP-MusicCaps~\cite{doh2023lp} and WavCaps~\cite{mei2023wavcaps} datasets, which contain instruction-input-output samples generated by GPT-3.5, we fine-tuned the LLaMA2-7B model using the Parameter-Efficient Fine-Tuning (PEFT) method, specifically Low-Rank Adaptation (LoRA)~\cite{hu2021lora}. LoRA freezes the pre-trained model weights and introduces trainable low-rank matrices into each layer. This approach enables the model to adapt to new data while retaining its foundational knowledge.

\begin{figure}[!t]
\centering
\includegraphics[width=0.92\linewidth]{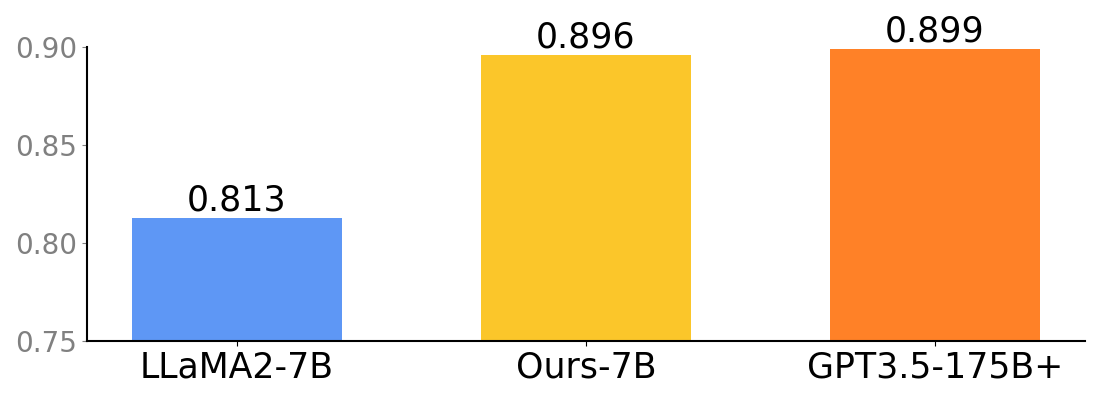}
\vspace{-4mm}
\caption{Tag-to-Caption Generation Results for LLaMA-7B, Finetuned LLaMA-7B (Ours), and GPT3.5-175B+.}
\vspace{-4mm}
\label{fig:k2c}
\end{figure}

Following previous work~\cite{doh2023lp}, we evaluate the caption generation model using the MusicCaps dataset, which includes tag lists and caption annotations. We generate pseudo-captions by concatenating the instruction and tag lists as model input. In Figure~\ref{fig:k2c}, we compare each model by calculating the BERT Score~\cite{zhang2019bertscore} between the generated pseudo-captions and human captions. In the tag-to-caption task, our proposed finetuned LLaMA2 with 7 billion parameters and demonstrates comparable performance to GPT3.5, which has over 175 billion parameters (0.896 vs 0.899). We employ this finetuned LLaMA2 to generate pseudo-captions for MSD, Audioset-Music, Music4All, and FMA, which lack caption annotations.

\begin{figure}[!t]
\centering
\includegraphics[width=0.8\linewidth]{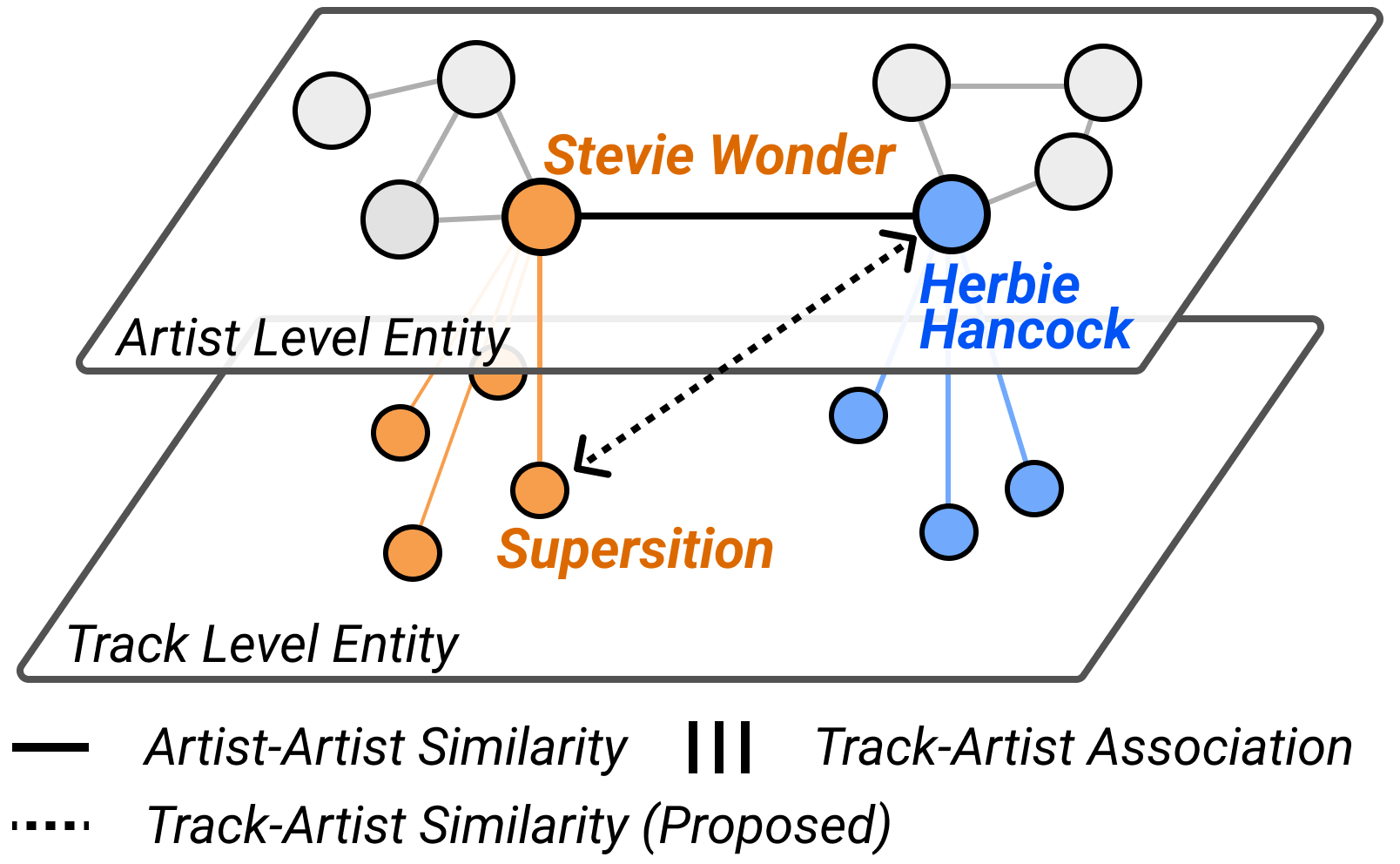}
\vspace{-3mm}
\caption{Connections between Artist and Track Entity.}
\vspace{-4mm}
\label{conncetion}
\end{figure}

\subsection{Music Description from Metadata Knowledge Graph}
\label{sec:metadata}
In the context of the query-by-metadata scenario, we construct metadata descriptions for track audio by leveraging music knowledge graphs. First, utilizing objective metadata associations and templates\footnote{music track \textit{\{title\}} by \textit{\{artist\}} from \textit{\{album\}}}, we create descriptions for track titles, artist names, and album names, paired with the corresponding track audio. Second, we propose the track-artist similarity, achieved by linking artist similarity connections and metadata associations (see Figure~\ref{conncetion}). For example, when we have information indicating the similarity between artists Stevie Wonder and Herbie Hancock, we inherit the artist similarity to Stevie Wonder's track Superstition, forming track-to-artist similarity with Herbie Hancock. The track-artist similarity is synthesized into a textual sample with the template\footnote{similar with artist \textit{\{artist\}}}. We leverage 14k artists from the training split of the OLGA dataset~\cite{korzeniowski2021artist} to establish a total of 8.9 million track-to-artist similarity connections for 552k MSD track items. During each training step, we randomly sample one similar artist to associate it with a track.

\subsection{Unified Joint Training Framework}
\label{sec:joint}
We employ a joint training approach that leverages a diverse range of datasets~\cite{gardner2021mt3,wu2022large}. This approach allows a single model to understand shared vocabulary semantics simultaneously while capturing the relationships between audio and various forms of natural language semantics (i.e., content, metadata, and relative similarity). Including distinct datasets within a single batch allows audio samples to be trained alongside more diverse negative texts. At the same time, textual data can be learned from a broader range of audio recording qualities and timbres. To mitigate the limitations of the low-resource dataset, we employ an explicit sampling proportion.

\section{Expriments}
\vspace{-1mm}
\label{sec:expriments}

\begin{table}[!t]
\resizebox{\linewidth}{!}{%
\begin{tabular}{lcccrrr}
\toprule
Dataset          & C         & M         & S         & \# item & Duration (h) & Prob. \\ \midrule
MSD              & \checkmark & \checkmark & \checkmark & 740.3k  & 6169.3       & 0.40   \\
Audioset (Music) & \checkmark &           &           & 381.3k  & 1059.0       & 0.25  \\
Music4All        & \checkmark & \checkmark &           & 86.4k   & 720.1        & 0.25  \\
FMA        & \checkmark &           &           & 92.4k   & 769.6        & 0.09  \\
MusicCaps        & \checkmark &           &           & 2.6k    & 7.4          & 0.01  \\ \midrule
Total            & \checkmark & \checkmark & \checkmark & 1.3M    & 8725.5       & 1.00   
\\ \bottomrule
\end{tabular}
}
\vspace{-3mm}
\caption{\textbf{Pre-training data}. \textbf{C} stands for the content annotation (e.g, tag, caption). \textbf{M} stands for objective metadata  (e.g, title, artist, album). \textbf{S} stands for relative similarity connection (e.g, similar artist).}
\vspace{-5mm}
\label{tab:data}
\end{table}


\subsection{Datasets}
\label{sec:datasets}

Our training dataset is a mixture of several existing tagging datasets, reported in Table~\ref{tab:data}. Except for the MSD dataset, we did not include evaluation split data in the training process. In addressing potential data leakage~\cite{weck2023data}, we conducted audio fingerprinting~\cite{ellis20142014} tests on the pre-training dataset between testsets (MusicCaps-eval~\cite{agostinelli2023musiclm}, Song Describer~\cite{manco2023song}) and systematically removed overlapping outcomes.

\vspace{1mm}
\noindent\textbf{MSD}: Million Song Dataset (MSD)~\cite{bertin2011million} is a collection of metadata and audio clips for 1 million tracks. In our research, we utilized two sources for content annotation: the ECALS subset~\cite{doh2023toward}, enriched with tags from Last.fm and AllMusic. Additionally, we leveraged artist similarity from the training split of OLGA dataset~\cite{korzeniowski2021artist} to construct a knowledge graph (described in Section~\ref{sec:metadata}).

\vspace{1mm}
\noindent\textbf{Audioset (Music)}: Audioset~\cite{gemmeke2017audio} is a collection of over 2 million 10-second audio clips excised from YouTube videos and labeled with 527 labels. We used music subset~\cite{huang2023noise2music}, which contains 381k clips with 130 music-related labels.

\vspace{1mm}
\noindent\textbf{FMA}: Free Music Archive (FMA)~\cite{defferrard2016fma} is a large-scale dataset with a rich set of track, artist, tag annotation. We use the 92k tracks and 161 genre tags from the training subset of FMA large.

\vspace{1mm}
\noindent\textbf{Music4All}: Music4All~\cite{santana2020music4all} includes 30-second
audio clips, lyrics, and 16 other metadata such as title, album name, and artist name of 109k tracks. Following data split from \cite{choi2021listen}, we used 86k tracks and 20k unique tags.

\vspace{1mm}
\noindent\textbf{MusicCaps}: MusicCaps~\cite{agostinelli2023musiclm} contains 5521 music examples, each labeled with a 13k unique aspect list and a free text caption written by musicians. We used 2663 samples from the training split.

\begin{table*}[!t]
\centering
\resizebox{\textwidth}{!}{%
\begin{tabular}{lclccccccccc}
\toprule
& Pretrain data &  & \multicolumn{2}{c}{MusicCaps~\cite{agostinelli2023musiclm}}   &  & \multicolumn{2}{c}{Song~Describer~\cite{manco2023song}} &  & DimSim~\cite{lee2020disentangled}       &  & OLGA~\cite{korzeniowski2021artist}           \\ \cmidrule{2-2} \cmidrule{4-5} \cmidrule{7-8} \cmidrule{10-10} \cmidrule{12-12} 
Model                          & Size / Hours  &  & Caption        & Tag            &  & Caption         & Tag             &  & Track          &  & Artist         \\ \midrule
Random                         & - / -           &  & 0.5          & 0.500          &  & 1.9           & 0.497           &  & 0.500          &  & 0.004          \\
CLAP-Fusion~\cite{wu2022large}                  & 2.6M / 4789   &  & 17.6          & 0.629          &  &  25.7               & 0.632           &  & 0.590               &  & 0.008          \\
TTMR~\cite{doh2023toward}                   & 0.5M / 4308   &  & 2.7          & 0.653          &  & 17.4           & 0.733           &  & 0.563          &  & 0.009          \\ \midrule
\textbf{TTMR++ (Ours)}                  &               &  &                &                &  &                 &                 &  &                &  &                \\
Baseline (Tag Concat.)         & 1.1M / 6879   &  & 13.6          & 0.686          &  & 32.7           & 0.739           &  & 0.570          &  &  0.010              \\
\hspace{2mm} + LLM Augmented Caption        & 1.1M / 6879   &  & 27.9          & 0.700          &  & 35.1           & 0.763           &  & 0.548          &  & 0.012          \\
\hspace{2mm} + Title, Artist, Album Text    & 1.3M / 8726   &  & 28.7 & 0.704          &  & \textbf{38.9}  & 0.762           &  & 0.647          &  & 0.076          \\
\hspace{2mm} + Track-to-Artist Similarity Text & 1.3M / 8726   &  & \textbf{29.2}          & \textbf{0.714} &  & 38.3           & \textbf{0.768}  &  & \textbf{0.669} &  & \textbf{0.187} \\ \bottomrule
\end{tabular}
}
\vspace{-3mm}
\caption{Results of text-to-music retrieval tasks, compared with baseline models. In the case of the proposed \textbf{TTMR++}, each row accumulates improvements from the rows above. In the last two rows, the data size expands as more audio samples are linked with metadata annotation.}
\vspace{-5mm}
\label{tab:results}
\end{table*}

\subsection{Evaluation}
\label{sec:eval}
We selected various datasets covering diverse textual query types, including caption, tag, track title, and artist name. For caption-based retrieval tasks, we present Recall@10 (R@10) metrics across two distinct datasets: MusicCaps-eval~\cite{agostinelli2023musiclm} and Song Describer~\cite{manco2023song}. 
In tag-based retrieval tasks, we report the ROCAUC on datasets identical to those employed in caption-based retrieval. To address challenges arising from numerous synonyms, we employ SentenceBERT~\cite{reimers2019sentence-bert} to consolidate tags with a similarity surpassing 0.9, utilizing only those tags annotated on a minimum of ten tracks.

Concerning track-based retrieval, we employed the clean subset of the dim-sim dataset~\cite{lee2020disentangled}, which comprises a curated set of music similarity triplets confirmed by human raters. Thanks to the similarity triplets, track-based retrieval becomes a task of finding music items more similar to a given text query rather than identifying them. To formulate text queries, we utilized metadata and templates\footnote{similar music track with \textit{\{title\}} by \textit{\{artist\}} from \textit{\{album\}}} from the anchor track. Our model extracts the text feature of anchors and audio features of two candidates, enabling us to measure similarity in joint embedding space. We report triplet prediction accuracy.

In artist-based retrieval, our evaluation relies on the OLGA dataset~\cite{korzeniowski2021artist}, which provides ground-truth cultural artist similarity data sourced from AllMusic. In alignment with prior work~\cite{ferraro2023contrastive}, we established the artist prototype vector by computing the average of track audio vectors derived from the MSD dataset. To formulate queries, we represented the query using the artist's name along with a template\footnote{similar artist with \textit{\{artist\}}}, presenting it as textual input. Within the joint embedding space, we calculated the similarity between the query artist text and the candidate artist audio prototype as the prediction. We report nDCG@200 to measure how accurately the joint embedding captures artist similarity.

\subsection{Implementation Details}
For all the experiments, the input of the encoder is a 10-second audio signal at 22050Hz sampling rate. It is converted to a log-scaled mel spectrogram with 128 mel bins, 1024-point FFT with a Hann window, and a hop size of 10~ms. All models are optimized using AdamW with a learning rate of 5e-5. The temperature parameter $\tau$ is initialized to 0.1 for all models. We used a cosine learning rate decay to zero after a warmup over 5000 steps. For the training, we used 768 batch-size and the models are trained for 32768 updates. To manage multiple paired text inputs such as tags, captions, and metadata, we apply a dynamic text dropout approach. Following \cite{doh2023toward, wu2023clamp}, for a given text with $L$ candidates, randomly sample $K$ texts. Here, $K$ is uniformly and randomly chosen from integers within the range of 1 to $L$. These selected texts are then concatenated in random order to create a single input text for the text encoder. 

\section{Results}
\label{sec:results}
The upper part of Table~\ref{tab:results} shows the text-to-music retrieval results with state-of-the-art dual encoder models: CLAP-Fusion~\cite{wu2022large}, a model trained on 2.6M general audio clips with feature fusion, and TTMR~\cite{doh2023toward}, a model trained on MSD audio with corrupted tag concatenation. In both tag-based and caption-based retrieval, both models exhibit strong performance. CLAP-Fusion~\cite{wu2022large}, which utilizes keyword-to-caption augmentation with the T5-based model, excels in caption-based retrieval, while TTMR, employing tag supervision, demonstrates its strength in tag-based retrieval. However, it is noteworthy that both models show lower performance in track-based and artist-based retrieval.

The lower part of Table~\ref{tab:results} presents the results obtained with the proposed TTMR++. The baseline model represents an extension of TTMR, incorporating elements such as joint training, increased batch size, and larger parameter models. We observe improvements in performance across all tasks, validating the importance of data size and semantic diversity from joint training. Next, we apply LLM augmented caption to the training datasets described in Section~\ref{sec:llm}. 
We observe significant improvements in caption-based retrieval with a marginal enhancement in tag-based retrieval. However, both models still exhibit relatively poor performance in metadata queries.


The last two rows of the Table~\ref{tab:results} depict the performance of models trained with metadata text, which includes titles, artist names, and album names. Upon incorporating metadata text, we observed significant performance improvements in track-based and artist-based retrieval. This indicates that the joint embedding model is capable of recognizing the characteristics of each artist or track by textual representations (title or name). Surprisingly, metadata text also leads to performance enhancements in caption-based retrieval. We believe that when metadata text is linked with audio, music-specific metadata information alleviates the gap between audio and text representations in a joint embedding space.

Upon incorporating track-to-artist similarity text, we observed a slight increase in overall retrieval performance, except for the caption query. Particularly noteworthy was the substantial improvement observed in artist-based retrieval. This suggests that the utilization of the metadata knowledge graph has effectively bridged the gap between the two modalities by establishing more connections between metadata text and audio.

\section{Conclusion}
\label{sec:results}
We proposed an improved text-to-music retrieval with a finetuned large language model and metadata knowledge graph. The proposed methodology extends the scope of available text queries for users by leveraging provided tag annotations and objective metadata. We trained a cross-modal dual encoder model with multiple data sources and showed improved generalization across query types, including caption, tag, track, and artist. We conducted a systemic evaluation of the text-to-music retrieval task and show our \textbf{TTMR++} achieves state-of-the-art results.

Our research presents several future explorations. Firstly, our pertaining dataset lacks information on mid-level tempo or key. Secondly, it is worth considering that the contrastive learning-based discriminative loss may have limitations in effectively modeling the many-to-many relationship between music and language. Exploring pseudo-labels with task-specific models and probabilistic embedding spaces might offer promising directions for future research.

\vfill\pagebreak
\bibliographystyle{IEEEbib}
\bibliography{strings}

\begin{thebibliography}{10}

\bibitem{turnbull2008semantic}
Douglas Turnbull, Luke Barrington, David Torres, and Gert Lanckriet,
\newblock ``Semantic annotation and retrieval of music and sound effects,''
\newblock {\em IEEE TASLP}, 2008.

\bibitem{choi2019zero}
Jeong Choi, Jongpil Lee, Jiyoung Park, and Juhan Nam,
\newblock ``Zero-shot learning for audio-based music classification and tagging,''
\newblock in {\em ISMIR}, 2019.

\bibitem{won2021multimodal}
Minz Won, Sergio Oramas, Oriol Nieto, Fabien Gouyon, and Xavier Serra,
\newblock ``Multimodal metric learning for tag-based music retrieval,''
\newblock in {\em ICASSP}, 2021.

\bibitem{manco2022contrastive}
Ilaria Manco, Emmanouil Benetos, Elio Quinton, and Gy{\"o}rgy Fazekas,
\newblock ``Contrastive audio-language learning for music,''
\newblock in {\em ISMIR}, 2022.

\bibitem{huang2022mulan}
Qingqing Huang, Aren Jansen, Joonseok Lee, Ravi Ganti, Judith~Yue Li, and Daniel~PW Ellis,
\newblock ``{MuLan}: A joint embedding of music audio and natural language,''
\newblock in {\em ISMIR}, 2022.

\bibitem{doh2023toward}
SeungHeon Doh, Minz Won, Keunwoo Choi, and Juhan Nam,
\newblock ``Toward universal text-to-music retrieval,''
\newblock in {\em ICASSP}, 2023.

\bibitem{weck2023wikimute}
Benno Weck, Holger Kirchhoff, Peter Grosche, and Xavier Serra,
\newblock ``Wikimute: A web-sourced dataset of semantic descriptions for music audio,''
\newblock {\em arXiv preprint arXiv:2312.09207}, 2023.

\bibitem{wu2023clamp}
Shangda Wu, Dingyao Yu, Xu~Tan, and Maosong Sun,
\newblock ``{CLaMP}: Contrastive language-music pre-training for cross-modal symbolic music information retrieval,''
\newblock in {\em ISMIR}, 2023.

\bibitem{gardner2021mt3}
Josh Gardner, Ian Simon, Ethan Manilow, Curtis Hawthorne, and Jesse Engel,
\newblock ``{MT3}: Multi-task multitrack music transcription,''
\newblock in {\em ICLR}, 2021.

\bibitem{wu2022large}
Yusong Wu, Ke~Chen, Tianyu Zhang, Yuchen Hui, Taylor Berg-Kirkpatrick, and Shlomo Dubnov,
\newblock ``Large-scale contrastive language-audio pretraining with feature fusion and keyword-to-caption augmentation,''
\newblock in {\em ICASSP}, 2023.

\bibitem{doh2023lp}
SeungHeon Doh, Keunwoo Choi, Jongpil Lee, and Juhan Nam,
\newblock ``{LP-MusicCaps}: Llm-based pseudo music captioning,''
\newblock in {\em ISMIR}, 2023.

\bibitem{mckee2023language}
Daniel McKee, Justin Salamon, Josef Sivic, and Bryan Russell,
\newblock ``Language-guided music recommendation for video via prompt analogies,''
\newblock in {\em CVPR}, 2023, pp. 14784--14793.

\bibitem{mei2023wavcaps}
Xinhao Mei, Chutong Meng, Haohe Liu, Qiuqiang Kong, Tom Ko, Chengqi Zhao, Mark~D Plumbley, Yuexian Zou, and Wenwu Wang,
\newblock ``Wav{C}aps: A {C}hat{GPT}-assisted weakly-labelled audio captioning dataset for audio-language multimodal research,''
\newblock {\em arXiv:2303.17395}, 2023.

\bibitem{touvron2023llama}
Hugo Touvron, Louis Martin, Kevin Stone, Peter Albert, Amjad Almahairi, Yasmine Babaei, et~al.,
\newblock ``Llama 2: Open foundation and fine-tuned chat models,''
\newblock {\em arXiv:2307.09288}, 2023.

\bibitem{radford2021learning}
Alec Radford, Jong~Wook Kim, Chris Hallacy, Aditya Ramesh, Gabriel Goh, Sandhini Agarwal, Girish Sastry, et~al.,
\newblock ``Learning transferable visual models from natural language supervision,''
\newblock in {\em ICML}, 2021.

\bibitem{oord2018representation}
Aaron van~den Oord, Yazhe Li, and Oriol Vinyals,
\newblock ``Representation learning with contrastive predictive coding,''
\newblock {\em arXiv:1807.03748}, 2018.

\bibitem{liu2019roberta}
Yinhan Liu, Myle Ott, Naman Goyal, Jingfei Du, Mandar Joshi, Danqi Chen, Omer Levy, Mike Lewis, Luke Zettlemoyer, and Veselin Stoyanov,
\newblock ``Roberta: A robustly optimized bert pretraining approach,''
\newblock {\em arXiv:1907.11692}, 2019.

\bibitem{ouyang2022training}
Long Ouyang, Jeffrey Wu, Xu~Jiang, Diogo Almeida, Carroll Wainwright, Pamela Mishkin, Chong Zhang, et~al.,
\newblock ``Training language models to follow instructions with human feedback,''
\newblock in {\em NeurIPS}, 2022.

\bibitem{taori2023stanford}
Rohan Taori, Ishaan Gulrajani, Tianyi Zhang, et~al.,
\newblock ``Stanford alpaca: An instruction-following llama model,''
\newblock in {\em https://crfm.stanford.edu/2023/03/13/alpaca.html}, 2023.

\bibitem{hu2021lora}
Edward~J Hu, Yelong Shen, Phillip Wallis, Zeyuan Allen-Zhu, Yuanzhi Li, Shean Wang, Lu~Wang, and Weizhu Chen,
\newblock ``Lora: Low-rank adaptation of large language models,''
\newblock {\em arXiv:2106.09685}, 2021.

\bibitem{zhang2019bertscore}
Tianyi Zhang, Varsha Kishore, Felix Wu, Kilian~Q Weinberger, and Yoav Artzi,
\newblock ``Bertscore: Evaluating text generation with bert,''
\newblock in {\em ICLR}, 2020.

\bibitem{korzeniowski2021artist}
Filip Korzeniowski, Sergio Oramas, and Fabien Gouyon,
\newblock ``Artist similarity with graph neural networks,''
\newblock in {\em ISMIR}, 2021.

\bibitem{weck2023data}
Benno Weck and Xavier Serra,
\newblock ``Data leakage in cross-modal retrieval training: A case study,''
\newblock in {\em ICASSP}, 2023.

\bibitem{ellis20142014}
Daniel Ellis,
\newblock ``The 2014 labrosa audio fingerprint system,''
\newblock in {\em ISMIR}, 2014.

\bibitem{agostinelli2023musiclm}
Andrea Agostinelli, Timo~I Denk, Zal{\'a}n Borsos, Jesse Engel, Mauro Verzetti, Antoine Caillon, Qingqing Huang, Aren Jansen, Adam Roberts, Marco Tagliasacchi, et~al.,
\newblock ``Music{LM}: Generating music from text,''
\newblock {\em arXiv:2301.11325}, 2023.

\bibitem{manco2023song}
Ilaria Manco, Benno Weck, Seungheon Doh, Minz Won, Yixiao Zhang, Dmitry Bodganov, Yusong Wu, Ke~Chen, Philip Tovstogan, Emmanouil Benetos, et~al.,
\newblock ``The song describer dataset: a corpus of audio captions for music-and-language evaluation,''
\newblock in {\em NeurIPS ML4Audio Workshop}, 2023.

\bibitem{bertin2011million}
Thierry Bertin-Mahieux, Daniel~PW Ellis, Brian Whitman, and Paul Lamere,
\newblock ``The million song dataset,''
\newblock in {\em ISMIR}, 2011.

\bibitem{gemmeke2017audio}
Jort~F Gemmeke, Daniel~PW Ellis, Dylan Freedman, Aren Jansen, Wade Lawrence, R~Channing Moore, Manoj Plakal, and Marvin Ritter,
\newblock ``Audio set: An ontology and human-labeled dataset for audio events,''
\newblock in {\em ICASSP}, 2017.

\bibitem{huang2023noise2music}
Qingqing Huang, Daniel~S Park, Tao Wang, Timo~I Denk, Andy Ly, Nanxin Chen, Zhengdong Zhang, Zhishuai Zhang, Jiahui Yu, Christian Frank, et~al.,
\newblock ``Noise2music: Text-conditioned music generation with diffusion models,''
\newblock {\em arXiv:2302.03917}, 2023.

\bibitem{defferrard2016fma}
Micha{\"e}l Defferrard, Kirell Benzi, Pierre Vandergheynst, and Xavier Bresson,
\newblock ``Fma: A dataset for music analysis,''
\newblock {\em ISMIR}, 2016.

\bibitem{santana2020music4all}
Igor Andr{\'e}~Pegoraro Santana, Fabio Pinhelli, Juliano Donini, Leonardo Catharin, Rafael~Biazus Mangolin, et~al.,
\newblock ``Music4all: A new music database and its applications,''
\newblock in {\em IWSSIP}, 2020.

\bibitem{choi2021listen}
Keunwoo Choi and Yuxuan Wang,
\newblock ``Listen, read, and identify: multimodal singing language identification of music,''
\newblock in {\em ISMIR}, 2021.

\bibitem{lee2020disentangled}
Jongpil Lee, Nicholas~J Bryan, Justin Salamon, Zeyu Jin, and Juhan Nam,
\newblock ``Disentangled multidimensional metric learning for music similarity,''
\newblock in {\em ICASSP}, 2020.

\bibitem{reimers2019sentence-bert}
Nils Reimers and Iryna Gurevych,
\newblock ``Sentence-{BERT}: Sentence embeddings using siamese bert-networks,''
\newblock in {\em EMNLP}, 2019.

\bibitem{ferraro2023contrastive}
Andres Ferraro, Jaehun Kim, Sergio Oramas, Andreas Ehmann, and Fabien Gouyon,
\newblock ``Contrastive learning for cross-modal artist retrieval,''
\newblock {\em arXiv:2308.06556}, 2023.

\end{thebibliography}

\vfill\pagebreak

\appendix

\end{document}